\documentclass[pre,onecolumn]{revtex4-2}
\usepackage{graphicx}
\usepackage{multibib}
\usepackage{natbib}

\begin{document}

\title{Effects of reduced gravity on the granular fluid-solid transition:
underexplored forces can dominate soft matter behaviors}

\date{\today}

\author{Paul S\'anchez}
\affiliation{CCAR, University of Colorado Boulder}
\author{Karen E. Daniels}
\affiliation{Deptartment of Physics, NC State University}
\author{Heinrich Jaeger}
\affiliation{Deptartment of Physics, James Franck Institute, U. Chicago}
\author{Troy Shinbrot}
\affiliation{Deptartment of Biomedical Engineering, Rutgers University}

\begin{abstract}
\bigskip
\centerline{\it Topical White Paper submitted to the NASA Biological and Physical Sciences in Space} 
\centerline{\it Decadal Survey 2023-2032}
\bigskip

Granular media are soft matter systems that exhibit some of the extreme behavior of complex fluids. Understanding of the natural formation of planetary bodies, landing on and exploring them, future engineering of structures beyond Earth and planetary defense all hinge on the ability to predict the complex mechanical behavior of granular matter. As we understand them, these behaviors are linked to the granular fluid to solid transition. In this white paper, we describe issues that emerge for granular systems under reduced gravity and their implications for basic science and space exploration.
\end{abstract}

\maketitle

\section{Introduction} 

\begin{figure}[b]
\hspace{0.1in} (a) \hspace{0.5in} (b)\hspace{1.2in} (c) \hspace{1.3in}(d)\hspace{1.4in} (e) \\
\includegraphics[width=1.12in,angle=90]{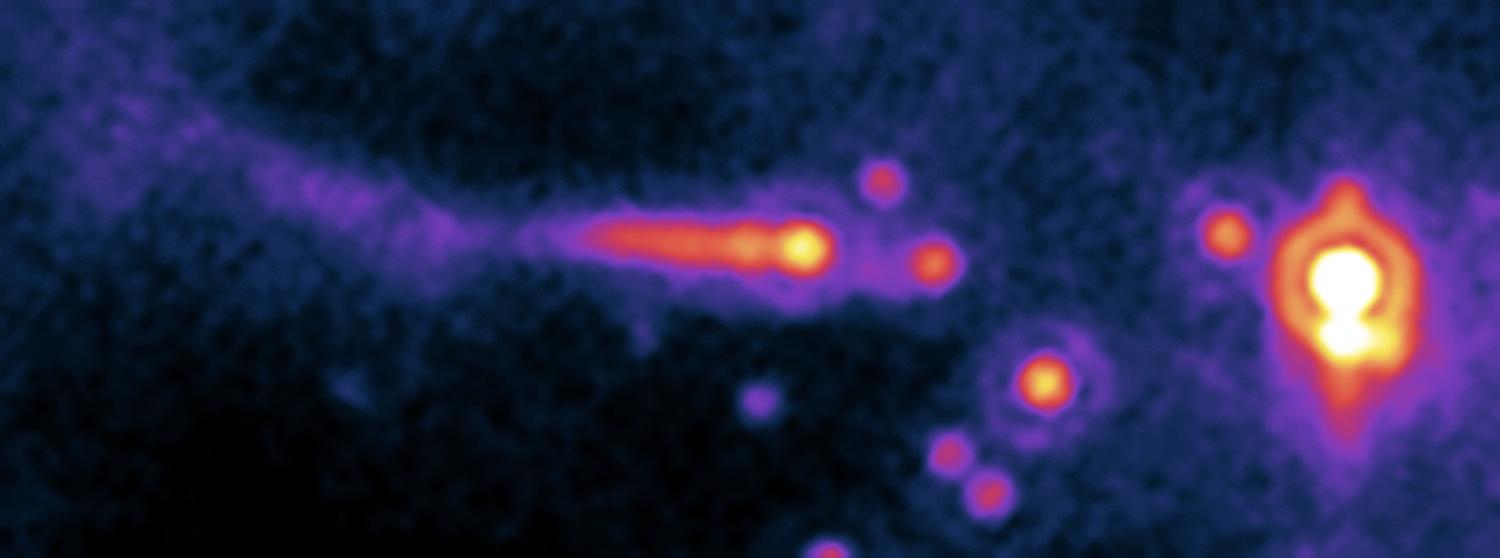} 
\includegraphics[height=1.12in]{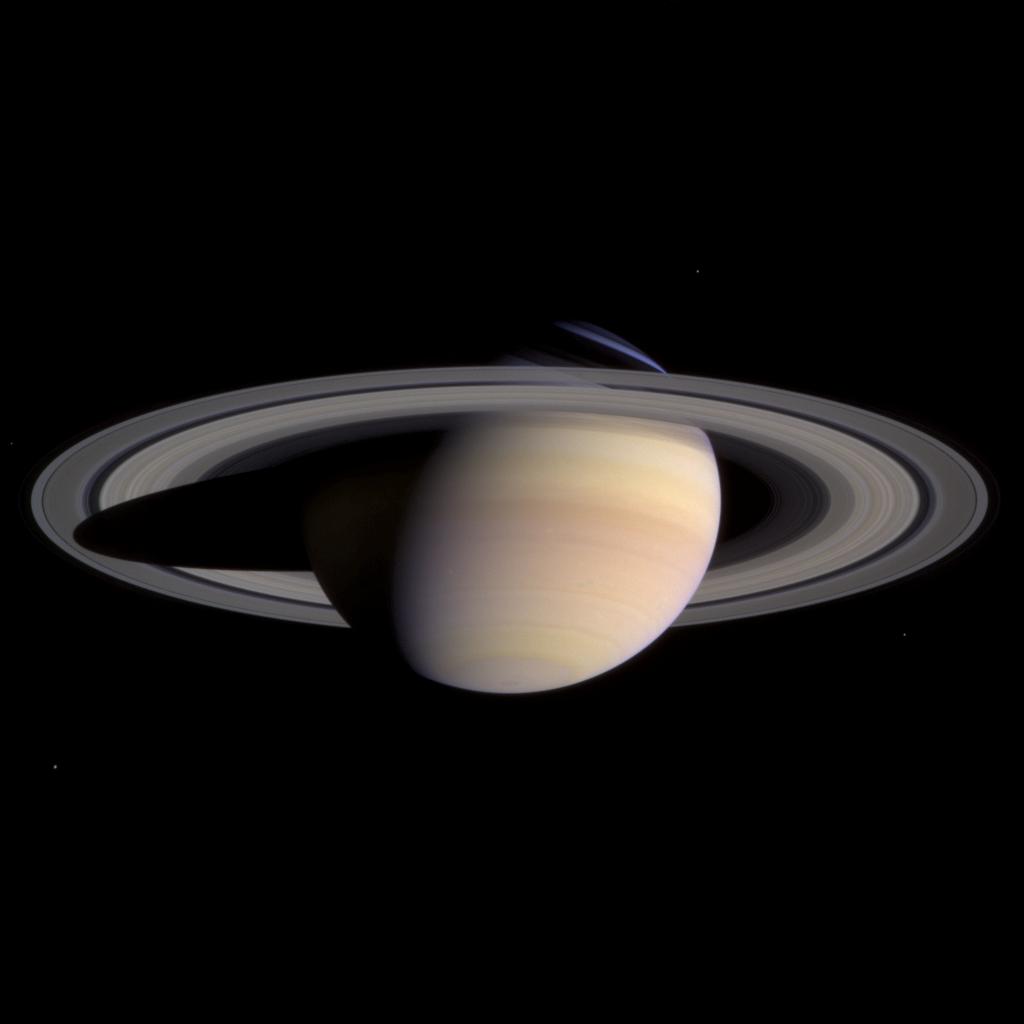} 
\includegraphics[height=1.12in]{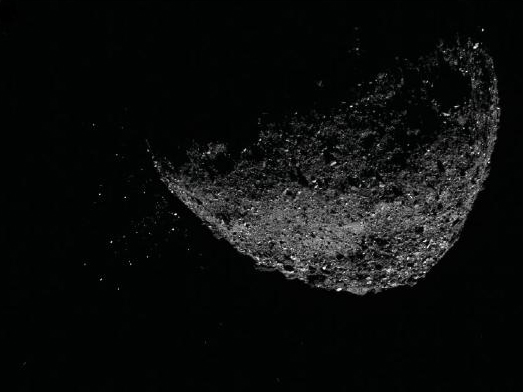} 
\includegraphics[height=1.12in]{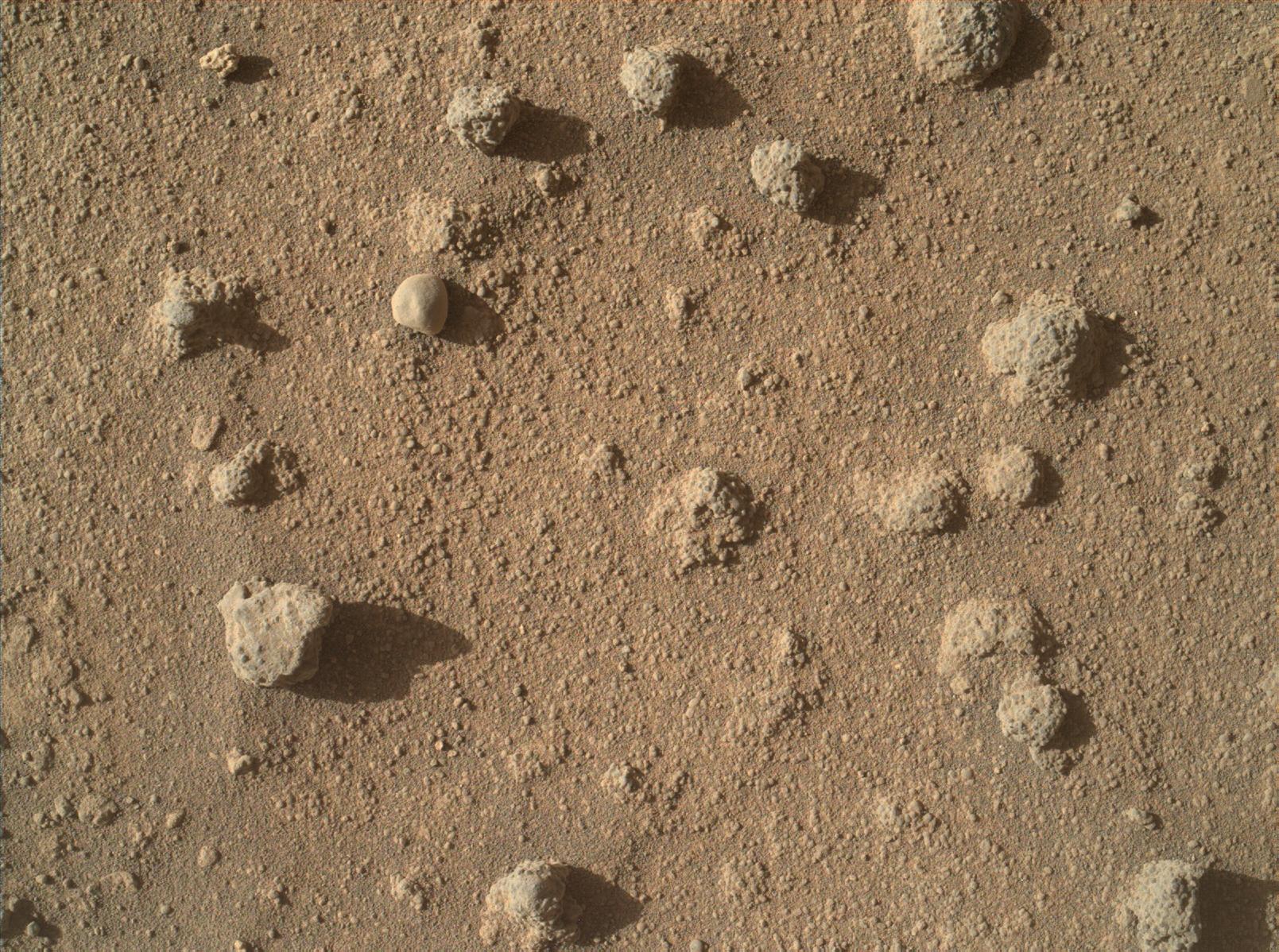}
\includegraphics[height=1.12in]{
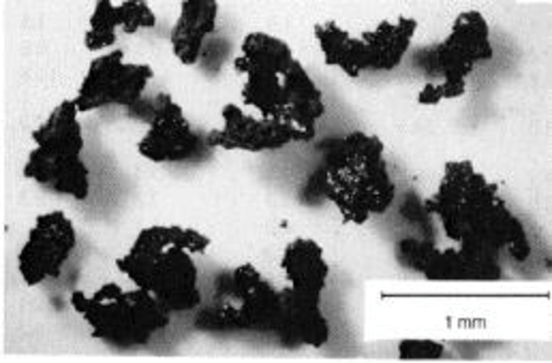} 
\caption{ {\bf Planetary granular materials, from large to small scale.}
(a) planet-forming disk being torn from a star, [Spitzer Space Telescope, NASA/JPL-Caltech/Univ.~of Ariz./Univ.~of Szeged]
(b) Saturn and its rings [Cassini, NASA/JPL/Space Science Institute]
(c) composite image of asteroid Bennu ejecting particles from its surface 
[NavCam 1, OSIRIS-REx, NASA/Goddard/University of Arizona/Lockheed Martin] 
(d) nodules of cemented sand grains within Martian sandstone [Mars Rover Curiosity, NASA/JPL-Caltech/MSSS]
(e) typical lunar soil agglutinates [Apollo 11, Lunar Sourcebook]}
\label{montage}
\end{figure}

Almost 160 years ago, Victor Hugo asked ``How do we know that the creation of worlds is not determined by the fall of grains of sand?'' Little did he now that now, in this space age --- when travelling to distant worlds is not a work of fiction --- we would be struggling with this very question.
Evidence increasingly indicates that under conditions of reduced gravity, interactions that would otherwise be overlooked between granular materials dominate the essential physics.
From the earliest days of the solar system, grains of every size and shape have collided, bounced, broken and accumulated in a perpetual cycle to form the bodies that we now see and visit.  Fig.~\ref{montage} shows planetary granular materials at a variety of length scales.

The variety of observed granular behaviors spans from Earth-like scenarios, where relatively strong gravity predominates, to satellites and asteroids with lower surface gravity, to self-gravitation in protoplanetary disks and planetary rings. Granular dynamics in Earth's gravity are being well-investigated by researchers in the fields of \textit{soft matter} and \textit{complex fluids}, but that ongoing research indicates that electrostatic, cohesive, and other forces that can often be disregarded under terrestrial gravity may be essential to novel granular dynamics and result in otherwise unanticipated granular structures.

There are many examples of the interplay between these various effects. At the smallest scale, pervasive infiltration of irregular and abrasive lunar regolith into suits and equipment led Apollo astronaut John Young to put it that, ``Dust is the number one concern in returning to the moon.'' 
At the millimetric scale, early stages of planetary formation are inhibited by what has been called the bouncing barrier: aggregates are too large to be held together by van der Waals forces and are too small to be influenced by gravitational attraction.  At kilometric scales, 
non-spherical shapes emerge on small moons and asteroids, and at still larger scales, rapid rotation sculpts and disrupts modest-sized asteroids.  Even at the planetary scale, surface features on Mars appear to have been carved by the flow of liquid water despite seemingly prohibitively cold temperatures and low pressures.  Mechanisms  underlying these features remain unclear, but have been proposed to be due to dry granular flow at reduced gravity.  
At all of these length scales, 
reduced gravity affects the transition between fluid- and solid-like behaviors in granular materials which, in turn, produce documented peculiarities ranging from unexplained aggregation to curious surface features. Throughout the solar system, granular interactions lead to unexpected outcomes that merit further, intensive study.

We propose therefore that focusing attention on changes in granular dynamics that arise under reduced gravity will improve predictive understanding of the natural development of features ranging from sand pools on asteroids and razorbacks on planets to depositional flows and striations around impact craters and equatorial ridges on satellites.  We argue that experimentally validated numerical and analytic models of granular behaviours under reduced gravity will also provide new and important insights.  Advances can be foreseen in applications including formation and evolution of small bodies and planetary rings, mitigation strategies for sand deposition on photocells, strategies to travel in rovers or other vehicles without hazardous slipping, and developments in the use and processing of granular materials in mining and construction operations.

\section{Small forces with big implications \label{sec:neweffects}}

The adjectives {\it big} and {\it small} can only be applied in comparison to something else. Here, that is the gravitational weight of a grain, which varies considerably according to the planetary context in which it resides.
Note that we call a granular system in low gravity one where the gravitational field is smaller than that on Earth's surface, is provided mainly by a single body that is external to the studied system, and is unaffected by the latter.

\begin{figure}
\centering
\raisebox{1in}{(a)}
\includegraphics[height=2in]{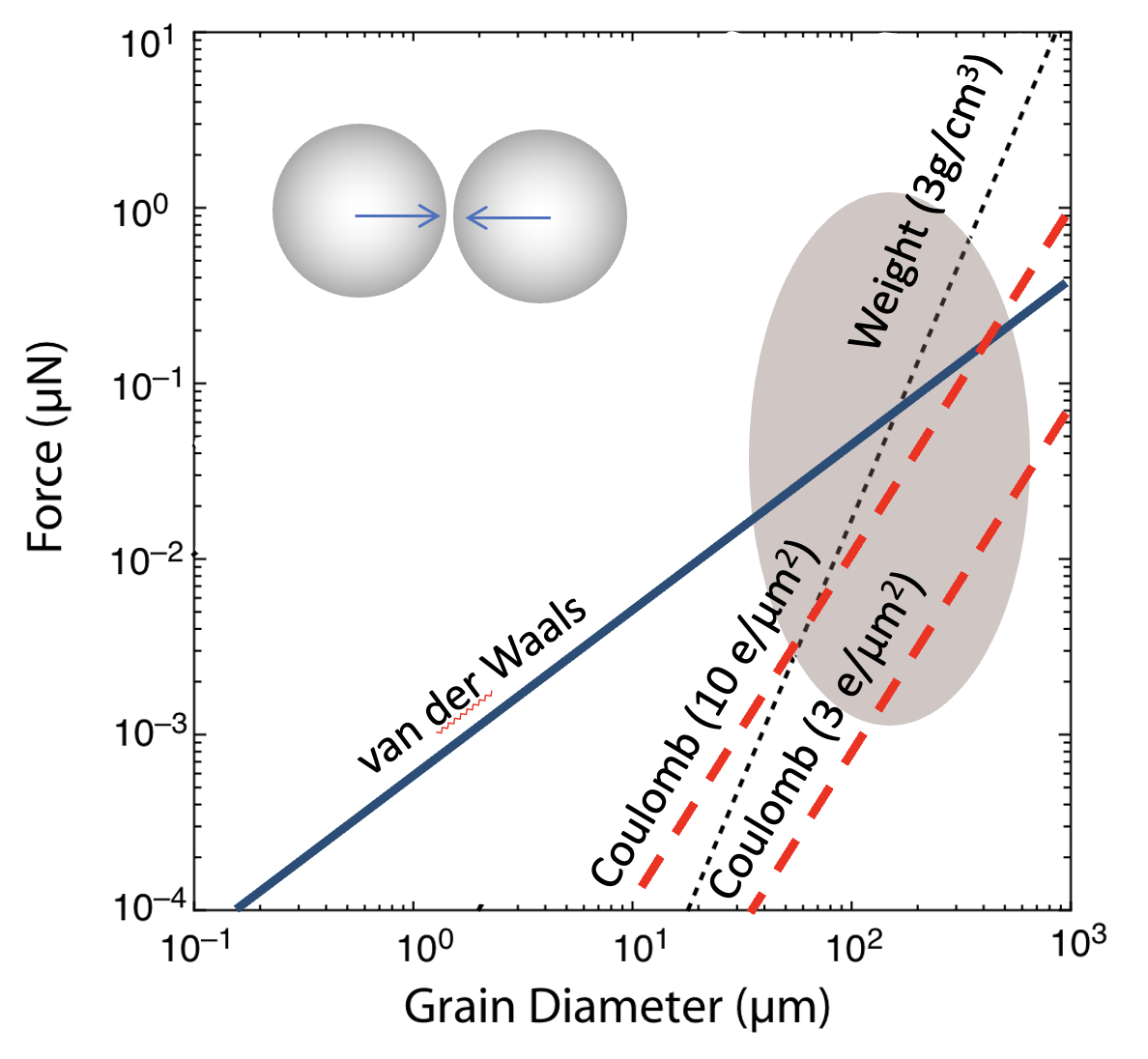} \hspace{1cm}
\raisebox{1in}{(b)}
\includegraphics[height=2in,trim={0 1cm 0 0},clip]{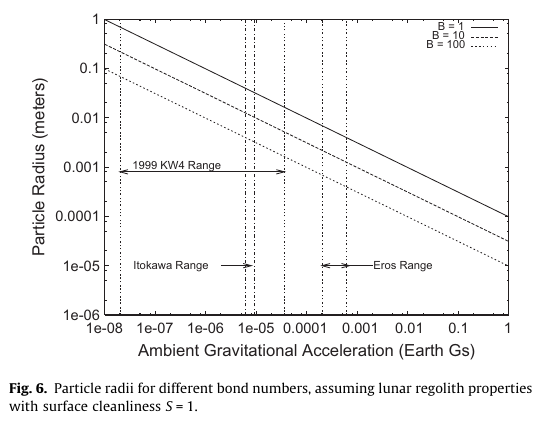}
\caption{(a) Comparison of the strength of van der Waals adhesion and  electrostatic (Coulomb) attraction for dielectric particles of various diameters in direct contact. The two curves for Coulomb forces correspond to mildly and more strongly charged insulating grains, with surface charge densities indicated in units of the electron charge $e$ = 1.6 x $10^{-19}$C. For comparison, the weight for an individual particle on Earth is plotted. Within the shaded region these various forces can lead to a complex interplay; for dust grains up to a few tens of microns short-ranged van der Waals forces dominate.
(b) Particle radii for different bond numbers, assuming lunar regolith properties with surface cleanliness of 1 (taken from \citet{scheeres2010}).}
\label{force_comparison}
\label{bondnumber}
\end{figure}

\paragraph*{Self-gravity: } 
In the absence of a nearby massive body, self-gravity arises from the tiny residual gravitational attraction between the many particles that form a granular system.
These forces can still have a measurable effect if they act over a long enough time, with self-gravity driving both the disruption and accretion of small planetary bodies. Because it is a long-range interaction force, it governs how dust, pebbles, rocks and boulders interacted during the morphogenesis of the solar system. Current observations of planetary rings and asteroids  offer us a window to this ever-present process.

\paragraph*{Electrostatics:}   Even on Earth, the most basic aspects of how charge is generated on dielectric particles and what mechanism dominates under which conditions remain unclear.  As gravity is reduced, these conceptual shortcomings grow in significance, since  effects due to electrostatics play increasing roles in the formation of aggregates. 
Contact charging via particle collisions is one of the main mechanisms that have been proposed for charge transfer between particles, but even the nature of the carrier of the  charge transferred (electrons, ions or bulk matter) is still debated \citep{shinbrot2008,lacks2019,waitukaitis2014,lee2018}. For ions, the particle surface chemistry is important and it must change with atmospheric composition. While it has been argued that H$^+$ and OH$^-$ ions dominate contact charging on Earth, due to residual, molecularly thin layers of water on most terrestrial bodies, Mars or smaller bodies largely lack adsorbed liquid water.  How materials develop charge and where charged particles stick are of manifest importance for preventing buildup on solar cells, machinery, and space suit joints.  

Recent work has shown that electrostatic forces can provide a path to overcoming the bouncing barrier, which prevents planetesimals from growing beyond pebble size: electrostatic forces can trap particles after a bounce, leading to repeated inelastic collisions and eventual sticking \citep{lee2015,steinpilz2020}. Additionally, the interaction of very small particles with the solar wind will give raise to what has been termed dusty plasmas, highly charged particle assemblies interacting via Coulomb forces.  Still, how particles acquire the required charges or charge distributions remains a major question. Symmetry notwithstanding, particles cannot all acquire the same charges, or else they would repel rather than stick. Thus, aggregation into ensembles requires complex charged micro-structures that are just beginning to be explored systematically \citep{lacks2019,lee2015,qin2016}.

\paragraph*{Cohesion: }  
Two particles may experience an attractive contact force either via cohesive (van der Waals) forces between their two surfaces, or mediated by small liquid bridges (Laplace pressure, via the surface tension of the liquid). Under Earth's gravity, such forces are often negligible for macroscopic particles, but in low-gravity environments their relevance can extend to larger size particles (see Fig.~\ref{bondnumber}), where contact forces can easily exceed the weight of a particle.  This effect is quantified by the Bond number, the ratio of the cohesive force between particles and their weight due to the gravitational attraction of other bodies.  This leads to distinctive effects such as the formation of fragile networks of grains that are held together chiefly by cohesion but are subject to collapse when disturbed. 

Fragile networks ({\it fairy castles}) are more likely to form in loose aggregates of cohesive grains under low gravity: a concern for mission-critical exploration and material processing.

\paragraph*{Particle Shape:}  
Real granular matter is, of course, not comprised of perfectly spherical particles.  Indeed, non-sphericity and non-concavity strongly affect collisional dynamics and thus the way particles can flow past each other, pack and form aggregates.  This in turn affects the conditions under which a granular assembly will transition between liquid-like, flowing, and solid-like, rigid, behavior (the so-called jamming phase transition). 
Thus, the ability to predict how a granular medium will respond to and transmit forces, including through entanglement, depends crucially on understanding the role of particle shape \citep{jaeger2015,gravish_entangled_2012}. 
Improved understanding of the behavior of  non-spherical particle ensembles is important for impact dynamics and also for the handling of granular media during mining and excavation operations as well as for predicting how landing craft or rover wheels will perform.

\section{Why low-gravity granular physics matters}

The basic understanding of granular systems in low-gravity (or self-gravitating) is interesting in itself; however, outer space is a natural laboratory.  As such we will take some of those systems so that the effects of the interplay of these forces are better understood.  

\paragraph*{Morphogenesis:} 
Many aspects related to the genesis and evolution of bodies in our astronomical backyard are currently not well understood. 
Saturn's moons such as Pan, Atlas and dwarf planet Vesta teem with unusual morphological features. 
Near-Earth asteroids Itokawa and Eros have somehow developed sandy regions separated from their rocky substrates.  
Asteroids Ryugu and Bennu have striking, spinning-top profiles whose emergence has led to a healthy debate.
On Mars, features like razorbacks, gullies and channels, or multi-layered striations and flows surrounding craters add to the mystery.  All of these have developed through self-gravitation from rubble ranging in size from dust to boulders larger than houses.

\paragraph*{Spinning: }
One of the effects of solar radiation on a small planetary body is to provide it with a net torque; this has been termed the Yarkovsky-O'Keefe-Radzievskii\allowbreak-Paddack (YORP) effect. 
This torque  causes an angular acceleration and as the spin rate increases, the body will eventually deform and disrupt.   When and how this reshaping and disrupting processes take place however, are intrinsically linked to the internal structure and strength of the granular aggregate.

\paragraph*{Space exploration of large bodies} 
Successful crewed and robotic missions to the Moon and Mars will depend on developing semi-permanent structures such as antenna towers, solar panels, and crew habitats. All of these will require in-situ resource utilization (ISRU). This includes building foundations and/or anchors, using locally-gathered  materials in order to reduce the mass that arrives by rocket. Techniques for creating concrete-like substitutes from regolith will be an important component of making this possible. 
Because the relative importance of particle shape and electrostatic forces, compared to gravity, is higher, new challenges emerge. For example, lunar dust (see Fig.~\ref{montage}e) stuck to the astronauts' space suits and other equipment can cause difficulties. However, these same effects could also be harnessed for ISRU purposes. 

Robots and crewed vehicles will need to safely and successfully navigate on the surface, from landing to locomotion to sampling. A convenient classification  for the relevant flow regime is the inertial number 
$
I \equiv \frac{\dot{\gamma} d}{\sqrt{P/\rho}}
$
where the amount of confining pressure $P$ is set by the gravitational acceleration, strain rate $\dot{\gamma}$ is set by the type of interaction, and the particle size $d$ and density $\rho$ are specific to the regolith. While some desired activities are likely to be at high inertial number ($I \gg 1$), for instance arresting a lander or firing a probe), locomotion typically takes place in the intermediate  regime ($I \approx 1$), and digging/anchoring can be done quite slowly ($I \ll 1$).

\paragraph*{Space exploration of small bodies} 
There are additional challenges when exploring small planetary bodies --- comets, asteroids, the objects in the Kuiper Belt and the Oort cloud, small planetary satellites, Triton, Pluto, Charon, and interplanetary dust --- for which the surface gravity reaches down into the microgravity regime ($10^6$ times smaller than on Earth). All these missions have to interact with the surface of their target body or with granular material around them.  As such, there are five main aspects that space missions need to tackle: landing, roving, impacting, sampling, and anchoring. 
The gravitational potential energy is so low that a small craft, or even a rock, impacting  at cm/s speed could generate a small crater on a regolith covered surface, and escape velocites for ejecta can also be in the cm/s regime.  Controlling these aspects  requires detailed understanding of the transition from solid- to fluid-like behavior of a granular assembly.

\section{Recommendations} 

In view of the evidence that novel and poorly-elucidated effects arise in granular media under conditions of reduced gravity, significant research is called for.  
Granular matter may appear simple, but it involves exceedingly complex physics, including those arising from collective effects. It covers a  large range of size scales (microns to kilometers) and a subtle interplay among both long-range and short-range forces.  These short-range forces furthermore are typically highly nonlinear and often depend on the relative velocity of particles coming into contact. Standard thermodynamic approaches are often not applicable because particles are not subject to thermal fluctuations from a uniform heat bath and ergodicity cannot be assumed. 
Tackling this set of challenging problems requires a deep and clear understanding of the basic physics that governs the systems at hand.  With this in mind, we make the following recommendations.

\paragraph*{Experiments exploring particle-particle collisions and aggregation/fragmentation in the presence of both long- and short-range forces:} At reduced gravity, electrostatic (Coulomb-type) interaction start to compete with gravity even for large particle sizes. Both are long-range forces but charge-based interaction are significantly more complex than gravity.  At short range or contact, additional complications arise from the presence cohesive interactions, for example due to van der Waals forces. How this interplay affects the conditions for growth of or fragmentation of small particle agglomerates has been studied during the last decade, but much is still to be done. Systematic experiments on model systems that explore particle-particle collisions and aggregation/fragmentation in the presence of both long- and short-range forces are required to shed light on this.

\paragraph*{Experiments exploring when and how small dielectric particles can become charged:} A variety of mechanisms have been proposed, and detailed experiments are urgently needed to validate these ideas. Several of the mechanisms rely on charge transfer via particle-particle collisions: testing this stringently is a major challenge involving the development of experiments that are able to non-invasively measure  the type and amount of charge transferred in a single collision event.

\paragraph*{A general need for more and better Earth-based experimental platforms:} Reduced gravity experiments are often performed in planes, sounding rockets, or orbiting stations.  All of these are highly energy intensive, expensive, and not widely available to researchers,  even if there are direct links to a space agency or to a specific need of a space mission.  Additionally, particularly in view of the climate crisis, it seems appropriate to increase our reliance on lower-energy platforms which have the additional benefit of being more accessible to a larger number of researchers. 
Examples include cubesats and nanosats, drop towers \citep{waitukaitis2013,carter2019,lee2015}  with catapults to extend time aloft, Atwood machines to reduce gravity \citep{altshuler_settling_2014}, acoustic levitation \citep{lee2018,lim2019}, air-fluidized beds, centrifugal surface flows, and tilted airtable surfaces.  All of these approaches have yielded new insights, are easier to build on a constrained budget, are reusable, and have a minimal carbon footprint.  Moreover, these ground-based experiments permit the use of charging and charge measurement devices without restriction based on size, weight or portability, and  larger number of experiments can be carried out to improve statistics. Therefore, we would like to see physics researchers expand their use of these alternative approaches at reduced gravity that do not suffer from the mentioned drawbacks.

\paragraph*{Theory:} The inability to perform full scale experiments puts a particular emphasis on the need for improved theory.  This involves several specific areas that have already been described. The theory needs to provide a fundamental understanding of the interplay, and scaling of relevant charging  mechanisms with long-range gravitational, electrostatic and potentially also magnetostatic forces, effects of interlocking shapes that may produce fragile granular networks, and comparison between net and multipole charge interactions. This last item increasingly appears to play a central -- but poorly understood -- role in the formation and evolution of small asteroids and planetesimals as it directly affects their mechanical strength.

\paragraph*{Simulations:} Progress on theory depends on experimental and computational validation.  There is little question that current simulations can provide important insights and allow us to probe behaviors and extract information that experiments cannot touch.  At the same time, including all of the various forces into simulations that are relevant under reduced gravity conditions is beyond the current state of the art.
The next generation of simulation codes will have to implement particle characteristics and particle-particle interactions including: self-gravity, gravitational fields without spherical symmetry, variations in particle shape, localized particle charges, cohesion/adhesion, magnetostatics, electrostatics (including polarization), and low velocity impact physics. This all will need to occur in conjunction with new experimental platforms, so that individual elements of these codes can be validated or corrected.

\paragraph*{The Human Component:} Importantly, there is also a pressing need to foster interdisciplinary efforts among scientists in different fields of research and the author list on this White Paper is evidence of that.  At the moment there is but a handful of researchers who move between Granular Dynamics, Planetary Sciences, and Space Exploration. Much more cross-pollination is needed in order to make significant progress. Some of the recommendations --- such as more-accessible experimental platforms --- could additionally drive innovation and collaboration among students, the next generation of scientist and engineers who would be at the vanguard of the development of new knowledge. We need many sciences and scientists to study the Earth; studying other, smaller worlds is no different.

The recommendations we have made here are by no means exhaustive; however, they are what we believe will allow us to move forward in our understanding.

\nocite{scheeres2010,sanchez2014,hestroffer2019,sanchez2021,sanchez2021b,murdoch-ast4-2015,ballouz2021,lee2018,jaeger2015,carter2019,lim2019,lee2015,murphy2017,waitukaitis2013,waitukaitis2014,steinpilz2020,lacks2019,shinbrot2008,qin2016,smallbodies,dekleer2021,waitukaitis2013,carter2019,lee2015,altshuler_settling_2014,lee2018,lim2019,persson2021}

\nocite{jerolmack_viewing_2019,altshuler_settling_2014,bogdan_laboratory_2020, brisset_regolith_2020,shinbrot_size_2017,colwell_low_2003, featherstone_stick-slip_2021, ruppert_towers_2021, shrivastava_material_2020,kozlowski_stress_2021, metzger_experiments_2018, qian_principles_2015, brzinski_depth-dependent_2013}

\bibliographystyle{apsrev}
\bibliography{references,ked-nasa-decadal}

\end{document}